\documentclass[preprint,aps,prl]{revtex4-1}
\usepackage{graphicx}
\usepackage{hyperref}
\usepackage[T1]{fontenc}
\usepackage[utf8]{inputenc}
\usepackage{epstopdf}
\usepackage{amsmath}
\usepackage{amsfonts}
\usepackage{ulem}

\begin{document}

\title{Elastic continuum theory: Fully understanding of the twist-bend nematic phases}
\author{G. Barbero}
\affiliation{Department of Applied Science and Technology,
Politecnico di Torino, Corso Duca degli Abruzzi 24, 10129 Torino, Italy.}
\author{L. R. Evangelista} \email{Corresponding Author: lre@dfi.uem.br}, \author{M. P. Rosseto}
\affiliation{Departamento de F\'isica, Universidade Estadual de Maring\'a\\ Avenida Colombo, 5790-87020-900 Maring\'a, Paran\'a, Brazil}

\author{R. S. Zola}
\affiliation{Universidade Tecnol\'ogica Federal do Paran\'a, Campus Apucarana,\\
Rua Marc\'ilio Dias 635, 86812-460 Apucarana, Paran\'a, Brazil}
\author{I. Lelidis}
\affiliation{Solid State Section, Department  of Physics, University of Athens,\\
Panepistimiopolis, Zografos, Athens 157 84, Greece.}

\date{\today}

\begin{abstract}
 The twist-bend nematic phase, $N_{\rm TB}$,  may be viewed as a heliconical molecular arrangement in which the director $\bf n$ precesses uniformly about an extra director field, $\bf t$. It corresponds to a nematic ground state exhibiting nanoscale periodic modulation. To demonstrate the stability of this phase from the elastic point of view, a natural extension of the Frank elastic energy density is proposed. The elastic energy density is built in terms of the elements of symmetry of the new phase in which intervene the components of these director fields together with the usual Cartesian tensors. It is shown that the ground state corresponds to a deformed state for which $K_{22} > K_{33}$. When the elastic free energy is interpreted in analogy with the Landau theory, it is demonstrated the existence of a second order phase transition between the usual and the twist-bend nematic phase, driven by a new elastic parameter playing a role similar to the one of the main dielectric anisotropy of classical nematics and being closely related to the bulk compression modulus representing the pseudo-layers of twist-bend nematic phases. A phase transition and the value of the nanoscale pitch are predicted in accordance to experimental results. 
\end{abstract}

\maketitle

\begin{figure}[!htb]
\includegraphics[scale=0.20]{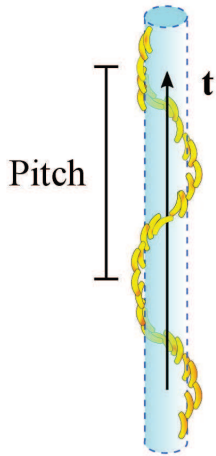}
\caption{One of the variants of the helical nematic phase in which the helix can wind  upwards or downwards. The vertical axis defines the direction of $\bf t$.}
\label{Figure_01}
\end{figure}

The recently discovered twist-bend nematic phase, $N_{\rm TB}$, has been theoretically predicted by Meyer~\cite{Meyer} and Dozov~\cite{Dozov}, and has been experimentally evidenced by a number of  
studies~\cite{Cest1,Chen1,Chen2,Sepelj,Imrie1,VPP1,Imrie2,Cest2,VPP2,VPP3,Beguin,Copic,Borshch}.
By speculating that the bend elastic constant, $K_{33}$, may become negative, Dozov~\cite{Dozov} employed a simple fourth-order model for the bulk free energy to predict the existence of two different periodic one-dimensional textures in the nematic phase. The first one was called splay-bend texture, characterized by a local bend that changes periodically its sign. The other one is a continuous conical twist-bend (TB) texture, in which the director $\bf n$ rotates along an axis forming a revolution cone with aperture $\theta$. In achiral system, this latter texture is twofold degenerated, allowing both right-hand and left-hand twists. By using techniques of small-angle x-ray scattering, modulated differential scanning calorimetry, it was recently concluded that the low-temperature mesophase of CB7CB is a new uniaxial nematic phase whose director distribution is composed of twist-bend deformations~\cite{Cest1}. Transmission electron (TEM) and polarized optical microscopy have been used to experimentally demonstrate the existence of the TB nematic phase~\cite{Borshch}. Sctructural observations confirming the existence of a TB ground state have been also carried out by TEM, together with measurements of the director cone angle and the full pitch of the director helix, indicating a strong coupling between the molecular and the director bend~\cite{Chen1}. These experimental identifications permitted to describe this nematic order as a true liquid-crystal phase with a new type of order. It is then a twist-bend modulated phase formed by achiral molecules, in which the director follows an oblique helicoid, maintaining a constant oblique angle $0 \le \theta \le \pi/2$,  with the helix axis~\cite{Borshch}. If the helix axis is along $z$, then the director $\bf n$ may be written as
$
{\bf n} = (\sin\theta \cos\varphi, \sin\theta \sin\varphi, \cos\theta),
$
where $\varphi$ is the azimuthal angle, given by
$\varphi = q z = (2\pi/p)z$,
where $q$ is the modulus of the wave vector and $p$ is the nanoscale pitch of the helix. 
In this way, this new ground state may be faced as an intermediate structural phase between the (usual) uniaxial and the chiral nematic phases~\cite{Borshch}. It can be also faced as a non-uniform ground state, characterized by the presence of a local spontaneous 
bend  and a very low value of the bend elastic constant $K_{33}$. 
To account for the elastic properties of the phase, Virga~\cite{Virga} proposed an intrinsically quadratic elastic theory to explain $N_{\rm TB}$ phases with an extra director field, in which there are two variants of the helical nematic phase, as predicted in Ref.~\cite{Dozov}, with an helix axis $\bf t$: one in which the helix winds upward and another one in which it winds downward, as shown in Fig.~\ref{Figure_01}. 
This theory does not require a negative bend elastic constant, as suggested by the quartic theory proposed in Ref.~\cite{Dozov} nor the existence of a locally ferroelectric phase, as required by the theory proposed in Refs.~\cite{Shamid1,Shamid2}, which is also quadratic but deals with an effective bend constant arising from the flexoelectric coupling. 

In this letter, a quadratic elastic theory also based on the existence of an extra director, represented by the helix axis $\bf t$, as in Ref.~\cite{Virga}, is proposed. Here, however, this extra director plays the role of an external field that couples with the nematic director $\bf n$. Differently from what is done in Ref.~\cite{Virga}, in which the $N_{\rm TB}$ is assumed as existing, the reference state used here is the undeformed one, characterized only by the uniform part of the free energy density. The aim is  to demonstrate that a phase like the twisted-nematic one may be stabilized, i.e., the ground state of the system corresponds to the minimum of a quadratic elastic theory built using only the existing elements of symmetry of the phase. To the best of our knowledge, this is the first theory capable of predicting the TB phase as a ground state, stable phase. It can also predict the TB to regular nematic phase transition and estimate parameters connected to experimental results.  Nevertheless, both theories reduce to the usual Frank's one for ordinary nematics if the extra director vanishes. 

The starting point is to consider a crystal characterized by a director ${\bf n}$ and a torsion ${\bf t}$ related to the chiral twisted collective arrangement. One assumes no polar order, in such a manner that ${\bf n}$ is the usual nematic director. The free energy of a crystal of this type, $f$, depends on the director field ${\bf n}$. According to general rules~\cite{Libro}
\vspace{-0.25truecm}
\begin{equation}
\label{1}
f({\bf n})=f_0({\bf n})+L_{ij}\, n_{i,j} +\frac{1}{2} K_{i,j,k,l} n_{i,j} n_{k,l}, 
\end{equation}
where the dots are terms of higher order in $n_{i,j}=\partial n_i/\partial x_j$, and the summation convention has been assumed. The quantity $f_0({\bf n})$, independent of $n_{i,j}$, can be decomposed, at the lowest order, as
\vspace{-0.25truecm}
\begin{equation}
f_0 ({\bf n}) = f_1 - \frac{1}{2} \eta \left({\bf n} \cdot {\bf t}\right)^2 + ..., 
\end{equation}
where $f_1$ is the uniform part of the energy of the usual nematic phase, if one treats the vector $\bf t$ as an external field, as proposed above. In this framework, the parameter $\eta$ plays the role of the main dielectric anisotropy of the nematic director with respect to $\bf t$~\cite{deGennes}. In this sense, $\eta$ may be viewed as the strength of the deformation of $\bf n$ about $\bf t$. It is then connected to the compression modulus of the pseudo-bulk layers B (as discussed in Ref.~\cite{RSC}).

The tensor of second order has to be decomposed in terms of the elements of symmetry of the phase, ${\bf n}$ and ${\bf t}$, of the identity tensor and of the antisymmetric tensor of elements $\delta_{ij}$ and $\varepsilon_{ijk}$, respectively.  Standard calculations give
\begin{equation}
\label{2}L_{ij}=A_1 n_i n_j+A_2 n_i t_j+A_3 t_i n_j+A_4 t_i t_j+A_5 \delta_{ij}+A_6 n_k \varepsilon_{kij}. 
\end{equation}
Since the medium is globally non-polar, the condition
\begin{equation}
\label{3}f({\bf n})=f(-{\bf n})
\end{equation}
holds. It requires that $L_{ij}$ has to be odd in ${\bf n}$, which implies that $A_1=A_4=A_5=0$. Furthermore, since $|{\bf n}|=1$, i.e  $n_i n_i=1$, it follows that $n_i n_{i,j}=0$. Consequently,  the term connected to $A_2$ does not play any role. The elastic energy density linear in the deformation, related to the spontaneous deformation of the phase under consideration, is
\begin{equation}
\label{4} L_{ij}\, n_{ij}=-\kappa_1 t_i\, n_j\, n_{i,j}+\kappa_2 n_k\,  \varepsilon_{kij}\, n_{i,j},
\end{equation}
where the phenomenological constants appearing into Eq.~(\ref{2}) have been renamed as $\kappa_1=-A_3$ and $\kappa_2=A_6$. A simple calculation shows that

\begin{equation}
\label{5}t_i\,n_j\, n_{i,j}=-{\bf t}\cdot [{\bf n}\times(\nabla \times {\bf n})]\quad{\rm and}\quad n_k \, \varepsilon_{kij} n_{i,j}={\bf n}\cdot (\nabla \times {\bf n}),
\end{equation}
and the linear term in the deformation of the elastic energy can be rewritten in covariant form as

\begin{equation}
\label{6}L_{ij}\,n_{i,j}=\kappa_1 \,{\bf t}\cdot [{\bf n}\times(\nabla \times {\bf n})]+\kappa_2\,{\bf n}\cdot (\nabla \times {\bf n}).
\end{equation}
One notices that while the term related to ${\bf n}\cdot (\nabla \times {\bf n})$ is a chiral term, present in standard cholesteric liquid crystals,  the first term is peculiar to the heliconical phase under consideration. In the case in which ${\bf t}={\bf u}_z$,  the director field ${\bf n}$ may be expressed as
\begin{equation}
\label{7}{\bf n}=[\cos \varphi(z)\, {\bf u}_x+\sin \varphi(z)\, {\bf u}_y)]\,\sin \theta+ \cos \theta\, {\bf u}_z,
\end{equation}
where ${\bf u}_x$, ${\bf u}_y$, ${\bf u}_z$ are the unit vectors along the Cartesian axes. As stated before, the director field given by Eq.~(\ref{7}) corresponds to a precession of ${\bf n}$ around $z$ with constant orientation of ${\bf n}$ with respect to ${\bf t}$.  In this case,  one easily shows that
\begin{equation}
\label{8}{\bf t}\cdot [{\bf n}\times(\nabla \times {\bf n})]=0 \quad {\rm and} \quad {\bf n}\cdot (\nabla \times {\bf n})=- q \sin^2 \theta
\end{equation}
are constant quantities.

\begin{figure}
\includegraphics[scale=0.25]{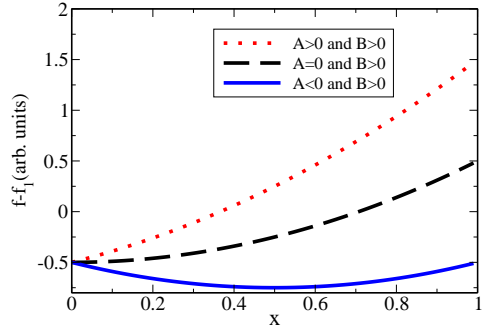}
\caption{Elastic energy density $F = f-f_1$ (in arbitrary units), given by Eq.~(\ref{14two}), as a function 
of $x$ in three situations: 1) for $A>0$ and $B>0$ (dotted line), 2) $A=0$ and $B>0$ (dashed line), and 3) $A<0$ and $B>0$ (solid line). For illustrative purposes, the curves have been drawn for $\eta >0$.  }
\label{Figure_02}
\end{figure}

The fourth-rank tensor,  $K_{ijkl}$, comparing in the elastic energy~(\ref{1}),  may be decomposed by following the standard procedure 
(see e.g., Ref.~\cite{Libro}), namely:
\vspace{-0.5truecm}
\begin{widetext}
\begin{eqnarray}
K_{ijkl} &=& k_5 n_jn_l\delta_{ik} + k_6 \delta_{ij}\delta_{kl} + k_7 \delta_{ik}\delta{jl} + k_8 \delta_{il}\delta_{jk} + p_1 n_ln_jt_it_k + q_1 t_it_jt_kt_l \nonumber \\
&+& \frac{1}{2}q_2 (t_it_j \delta_{kl} + t_k t_l \delta_{ij})
+ q_3 t_it_k\delta_{jl} + \frac{1}{2}q_4 (t_it_l\delta_{jk} + t_jt_k\delta_{il}) \nonumber \\
&+& q_5 t_jt_l\delta_{ik} + r_1 t_i \varepsilon_{jkl}.
\end{eqnarray}
\end{widetext}
The terms involving $k_i$ yield the usual Frank contribution. The other terms are also decomposed in the usual manner and the total elastic energy density may be written as:
\vspace{-0.5truecm}
\begin{widetext}
\begin{eqnarray}
f = f_0 &-& \frac{1}{2} \eta ( {\bf n} \cdot {\bf t})^2  + \kappa_1 \,{\bf t}\cdot \left[{\bf n}\times(\nabla \times {\bf n})\right]+\kappa_2\,{\bf n}\cdot (\nabla \times {\bf n}) + \frac{1}{2}K_{11} (\nabla \cdot {\bf n})^2
+ \frac{1}{2} K_{22} \left[{\bf n} \cdot (\nabla \times {\bf n})\right]^2
 \nonumber \\
&+& \frac{1}{2} K_{33} ({\bf n} \times \nabla \times {\bf n})^2 - (K_{22} + K_{24}) \nabla \cdot ({\bf n} \nabla \cdot {\bf n} + {\bf n} \times \nabla \times {\bf n}) \nonumber \\
&+& p_1 [{\bf t} \cdot ({\bf n} \times \nabla \times {\bf n})]^2 + q_1 [{\bf t} \cdot \nabla ({\bf t} \cdot {\bf n})]^2 + q_2 [{\bf t} \cdot \nabla ({\bf n} \cdot {\bf t}) (\nabla \cdot {\bf n})]\nonumber \\
&+& q_3 [\nabla ({\bf t}\cdot{\bf n})]^2 + q_4 [({\bf t} \cdot \nabla) {\bf n}]^2 + q_5 [\nabla ({\bf n} \cdot {\bf t})\cdot ({\bf t} \cdot \nabla) {\bf n}] - r_1 \nabla ({\bf n}\cdot{\bf t})\cdot (\nabla \times {\bf n}). \nonumber \\
\end{eqnarray}
\end{widetext}
\vspace{-0.5truecm}
For the physical state represented by the director given by Eq.~(\ref{7}), the terms connected with $q_1$,  $q_2$, $q_3$, $ q_5$, and $r_1$ do not contribute, because ${\bf n} \cdot {\bf t}=\cos\theta$, which is independent of $z$.
The only term that survives is the one connected with $q_4$, because:
$$
[({\bf t} \cdot \nabla) {\bf n}]^2 = q^2 \sin^2\theta.
$$
For what concerns the usual Frank terms, the non-vanishing ones are
$$
({\bf n} \times \nabla \times {\bf n}) = q^2 \sin^2\theta \cos^2\theta \quad {\rm and} \quad
({\bf n} \cdot \nabla \times {\bf n})^2 = q^2 \sin^4 \theta. 
$$
After collecting all the non-vanishing terms, one obtains for the elastic energy density the following expression:

\begin{eqnarray}
\label{14one}
f(q, x) &=& f_1 - \frac{1}{2} \eta (1-x) - \kappa_2 q x + \frac{1}{2} K_{22} q^2 x^2 \nonumber \\
&+& \frac{1}{2} K_{33} q^2 x (1-x) + q_4 q^2 x,
\end{eqnarray}
where $x= \sin^2\theta$. The procedure is now to minimize $f(q,x)$ with respect to $x$ and $q$. By imposing $\partial f/\partial x = 0$ and $\partial f/\partial q = 0$,
one obtains

\begin{equation}
\label{16one}
x^* = -\frac{K_{33} + 2 q_4 - \kappa_2 \sqrt{(K_{33} + 2q_4)/\eta}}{K_{22} - K_{33}}
\end{equation}
for the cone angle of $\bf n$ with $\bf t$, and 

\begin{equation}
\label{17one}
q^* = \sqrt{\frac{\eta}{K_{33} + 2 q_4}}
\end{equation}
for the modulus of the wave vector. The director profile associate to $x^*$ and $q^*$ corresponds to a minimum of the free energy density of an unlimited nematic sample only if
\begin{equation}
\label{18one}
\left( \frac{\partial^2 f}{\partial x^2}\right)_{x^*, q^*} \ge 0, 
\end{equation}
and the Hessian determinant defined by 
\begin{equation}
\label{19one}
H(x^*, q^*) = \left\{\frac{\partial^2 f}{\partial x^2}\frac{\partial^2 f}{\partial q^2} - \frac{\partial^2 f}{\partial x \partial q}\frac{\partial^2 f}{\partial q \partial x}\right\} > 0. 
\end{equation}
Simple calculations yield
\begin{equation}
\label{20one}
\left( \frac{\partial^2 f}{\partial x^2}\right)_{x^*, q^*} = \eta \frac{K_{22} - K_{33}}{K_{33} + 2 q_4} \ge 0
\end{equation}
and 
\begin{equation}
\label{21one}
H(x^*, q^*) = \kappa_2 \sqrt{\eta (K_{33} + 2 q_4)} - \eta (K_{33} + 2 q_4) >0.
\end{equation}
These are the requirements to be fulfilled in order to have an equilibrium state.
From Eqs.~(\ref{20one}) and~(\ref{21one}), one concludes that two scenarios are possible. Since $q^*$ is real, one obtains that a stable ground state may exist if $\eta <0$ and $K_{33}+ 2 q_4 <0$. In this case, $\eta (K_{22} - K_{33})/(K_{33} + 2 q_4) \ge 0$ when $K_{22} > K_{33}$. On the other hand, another stable ground state may be obtained if $\eta >0$ and $K_{33}+ 2 q_4 >0$. Again, $\eta (K_{22} - K_{33})/(K_{33} + 2 q_4) \ge 0$ when $K_{22} > K_{33}$. Thus, both situations are physically possible, implying that the twist elastic constant is such that $K_{22}> K_{33}$, as observed experimentally (see, e.g.,Ref.~\cite{Copic2}).  Now, the analysis can be cast in the framework of a simple Landau-like expansion of the free energy if one rewrites  Eq.~(\ref{14one}) in a still more compact form as
\begin{equation}
\label{14two}
f(x) = f_1 - \frac{\eta}{2} + A \, x + B\, x^2,
\end{equation}
in which $ A = -{\overline\kappa}_2 + \eta/2+  \overline{K_{33}}/2 $ and
$B = \Delta/2$,
with the following compact notation introduced: 
${\overline\kappa}_2 = \kappa_2 q$,  $\overline{K_{33}} = (K_{33} + 2 q_4) q^2$, 
and  $\Delta = (K_{22}-K_{33}) q^2$. Simple calculations show that $f$ is minimum for
$x = -A/( 2B)$, as given by (\ref{16one}) when (\ref{17one}) is taken into account. 
This corresponds to a stable state if $ B>0$, i.e., $\Delta >0$ or $K_{22} > K_{33}$. 
The twist bend phase, which corresponds to $x\ne 0$, is energetically favored only if $A<0$ 
(see Fig.~\ref{Figure_02}). Thus, there is a phase transition from the usual nematic phase ($x=0$) to a twist-bend nematic $x\ne 0$ for $A=0$, i.e., when $2\kappa_2 q = \eta +  \left(K_{33} + 2 q_4\right) q^*$, which, by using (\ref{17one}), becomes
$\eta = \eta_c=\kappa_2^2/ (K_{33} + 2 q_4)$.
For this critical value, the wavevector of the modulated phase assumes the value
$q_c = | \kappa_2/(K_{33} + 2 q_4)|$. In addition, $A = \eta - \sqrt{\eta \eta_c}$ may be 
used to drive the transition according to the values of $\eta$. Since the elastic energy density is being interpreted in the framework of a Landau theory, one can conclude that a second order phase transition occurs when $\eta=\eta_c$, if  $x=\sin^2\theta$ is used as the ``order parameter'' for the $N_{\rm TB}$ phase. This critical value of $\eta$ separates the usual from the twist-bend nematic phase. In this way as well, the parameter $\eta$ may play a role similar to the temperature such that $A>0$, for $\eta > \eta_c$, and 
$A <0$, for $\eta < \eta_c$. The order of magnitude of $\eta$ in the pure $N_{\rm TB}$ phase may be easily estimated if, for simplicity, one neglects the cholesteric contribution ($\kappa_2=0$), since the phase is not chiral. From Eq.~(\ref{16one}), one obtains $\eta = q^2 K_{22} \sin^2\theta - K_{33} q^2 \sin^2\theta$ or $\eta = 
B/\sin^2 \theta - (q^2 K_{33}/\sin^2\theta) \sin^4 \theta$, where $B = K_{22} q^2 \sin^4\theta$ is the compression modulus mentioned above~\cite{RSC}. Since $K_{22} > K_{33}$, one can assume that $\eta \approx B/\sin^2\theta$.  By using $B \approx 2\,$kPa, if, from~(\ref{17one}), we consider that $q = \sqrt{B/(K_{22}-K_{33}) \sin^2\theta}$, with $\theta = 16^{\circ}$, 
$K_{22} \approx 10^{-12}\,$N,  and $K_{33} \approx 10^{-13}\,$N,   one obtains
that $p_0 = 2\pi/q \approx 10\, $nm, in good agreement with experimental results for helix pitch of the twist-bend phase~\cite{Chen1,Borshch,RSC}. These estimations indicate that the new parameter not only can be expressed as a combination of known parameters characterizing the $N_{\rm TB}$ phase, but is also found to have an order of magnitude in good agreement with the available experimental results.

Some concluding remarks are in order. The experimentally discovered twist-bend nematic phase may be described from the elastic point of view by means of a quadratic theory,  which can be constructed as the usual one, i.e., by considering only the elements of  symmetry enough to characterize the system. For the twist-bend phase, these elements are the director, $\bf n$, and the helix vector $\bf t$, the latter playing a geometrical role similar to the one of an external field. In addition to the usual Frank constants, the elastic theory requires the existence of three parameters, namely, $\eta$, which plays a role similar to the one of the main dielectric anisotropy in usual nematics, $\kappa_2$, which is the coefficient of the twist distortion, peculiar to the heliconical phase,  and $q_4$, a parameter that renormalizes the elastic constant $K_{33}$.  The model can be interpreted in analogy with the Landau phenomenological approach for phase transitions. This analogy permits to treat the cone angle $\theta$ as the main ingredient for an "order parameter" defined by $x= \sin^2\theta$. The twist-nematic phase is then the one for which $x\ne 0$,  and is energetically favored with respect to the usual nematic phase ($x=0$) when $\eta = \eta_c$. There is a critical value of the parameter $\eta$ pointing a second order transition between the two stable phases allowed by symmetry considerations. This scenario is characterized by the exigence that $K_{22} > K_{33}$ always, as experimentally evidenced. In addition, the possibility of having negative values of $K_{33}$, as argued by some authors, may not be excluded but it is not mandatory. Indeed, one of the possible stable phases may exist for 
$\eta <0$ and $K_{33} + 2 q_4<0$, but this condition may be satisfied for $K_{33}>0$ when the elastic parameter $q_4$ is negative enough, i.e., when the term  
$q_4 [({\bf t} \cdot \nabla) {\bf n}]^2 = q_4 q^2 \sin^2\theta $ is energetically favored, which is a physically sound possibility. 
\section*{Acknowledgements} 
Many thanks to E. Virga (Italy) for helpful discussions and inspiring ideas. 
This work has been partially supported by the THALES Program No. 380170 of the Operational Project ``Education and Lifelong Learning'',
co-financed by the European Social Fund and the State of Greece, and by the INCT-FCx --National Institute of Science and Technology of Complex Fluids -- CNPq/FAPESP (Brazil).

\end{document}